# Infinite Networks of Identical Capacitors

**J. H. Asad[†], R. S. Hijjawi[††], A. J. Sakaji[†††] and J. M. Khalifeh[††††]**
[†] *Department of Physics, Tabuk University, P.O.Box.1144, Tabuk, Kingdom of Saudi Arabia.*
*E-Mail: jhasad1@yahoo.com.*
[††] *Department of Physics, Mutah University, Jordan.*
*E-Mail: Hijjawi@mutah.edu.jo.*
[†††] *Department of Physics, Ajman University, UAE.*
*E-Mail: a_sakaji@yahoo.com.*
[††††]*Dep. of Physics, Jordan University, 11942 Amman- Jordan*
*E-mail: jkalifa@ju.edu.jo.*

## Abstract

The capacitance between the origin and any other lattice site in an infinite square lattice of identical capacitors each of capacitance *C* is calculated. The method is generalized to infinite Simple Cubic (SC) lattice of identical capacitors each of capacitance *C*. We make use of the superposition principle and the symmetry of the infinite grid.

# 1. Introduction

A classic problem in electric circuit theory studied by many authors over many years is computation of the resistance between two nodes in a resistor network. Besides being a central problem in electric circuit theory, the computation of resistances is also relevant to a wide range of problems ranging from random walk [1,2], theory of harmonic functions [3] to first-passage processes [4] to Lattice Green's Functions [5] (LGF). The connection with these problems originates from the fact that electrical potentials on a grid are governed by the same difference equations as those occurring in the other problems. For this reason, the resistance problem is often studied from the point of view of solving the difference equations, which is most conveniently carried out for infinite networks. Kirchhoff [6] formulated the study of electric networks more than 150 years ago. The electric- circuit theory is discussed in detail in a classic text by Van der Pol and Bremmer [7] where they derived the resistance between nearby points on the square lattice.

In the previous 60 years many efforts were focused on analyzing infinite networks of identical resistors. In these efforts scientists used different methods. For example, Francis J. Bartis [8] introduced how complex systems can be treated at the elementary level and showed how to calculate the effective resistance between adjacent nodes of different lattices of one-ohm resistors. His note is interesting and educational. Venezian [9], Atkinson and Van Steenwijk [10] used the superposition of current distribution to calculate the effective resistance between adjacent sites on an infinite networks (i.e. Square, SC, Hexagonal,…). The mathematical problem involves the solution of an infinite set of linear, inhomogeneous difference equations which are solved by the method of separation of variables. Numerical results for the resistances between arbitrary sites are presented.
Monwhea Jeng [11] introduced a mapping between random walk problems and resistor network problems, where his method was used to calculate the effective resistance between any two sites in an infinite two-dimensional square lattice of unit resistors and the superposition principle was used to find the effective resistances on toroidal- and cylindrical-square- lattices. In the last decade many papers [12–18] have been published using an alternative method based on the LGF. The resistance between two arbitrary lattice sites for different infinite networks of identical resistors were studied for both the perfect and the perturbed networks. Finally, Wu [19] obtained the resistance between two arbitrary nodes in a resistor network in terms of the eigenvalues and eigenfunctions of the Laplacian matrix associated with the network. Explicit formulae for two point resistances are deduced in his paper for regular lattices in one, two, and three- dimensional under various boundary conditions.

Little attention has been paid to infinite networks consisting of identical capacitances $C$. Van Enk [20] studied the behavior of the impedance of a standard ladder network of capacitors and inductors where he analyzed it as a function of the size of the network. Recently, Asad et al [21-23] and Hijjawi et al [24] investigated many infinite lattices of identical capacitors using the LGF method and Charge distribution method. In these papers numerical results for the equivalent capacitance between the origin and any other lattice site was presented-using the above two methods- for the perfect infinite square network. Numerical results was also presented for the so-called perturbed infinite square network which results by removing one or two bonds from the perfect network.

In this paper we investigate analytically and numerically the capacitance between arbitrary lattice sites in an infinite square and SC grids using the charge distribution method which is based upon the superposition principle. The asymptotic behavior is also studied for large separation between the two sites. The basic approach used here is similar to that followed by Atkinson and Steenwijk [10].

Finally, it is important to mention that problems involving large or infinite electrical resistive networks are interesting and educational [25,26]. Applications include geophysical exploration with electrical currents, petroleum flow in oil wells, and random walks [27].

## 2. Infinite Square Lattice

In this section, we consider an infinite square network consisting of identical capacitances $C$. Let us define the voltage at the node $(l,m)$ to be given by $V_{l,m}$, and suppose that a charge of $Q_{l,m}$ enters that node from an external source.

Now using Ohm's and Kirchhoff's laws, we can write:

$$Q_{l,m} = (V_{l,m} - V_{l+1,m})C + (V_{l,m} - V_{l-1,m})C + (V_{l,m} - V_{l,m+1})C + (V_{l,m} - V_{l,m-1})C. \quad (2.1)$$

$$= 4V_{l,m}C - V_{l+1,m}C - V_{l-1,m}C - V_{l,m+1}C - V_{l,m-1}C. \quad (2.2)$$

We shall look for an integral representation for $V_{l,m}$, and take it to be in the form:

$$V_{l,m} = \int_0^{2\pi} d\beta F(\beta) V_{l,m}(\beta). \quad (2.3)$$

with
$$V_{l,m}(\beta) = \exp(i|l|\alpha + im\beta). \tag{2.4}$$
Here $\alpha$ is a function of $\beta$.

The above representation is a modified Fourier transform.
For $l \succ 0,$ we get:

$$4V_{l,m}(\beta) - V_{l+1,m}(\beta) - V_{l,m+1}(\beta) - V_{l,m-1}(\beta) = 2\exp(il\alpha + im\beta)[2 - Cos\alpha - Cos\beta]. \tag{2.5}$$

From the above equation, we require (i.e. in order that the contributions of the potential vanishes) $\alpha$ to be related to $\beta$ as:

$$Cos\alpha + Cos\beta = 2. \tag{2.6}$$

In a similar way, we find zero for this contribution if $l \prec 0$. Thus, for any integrable $F(\beta)$, then $Q_{l,m}$ given in Eq. (2.1) goes to zero, unless $l = 0$.
For $0 \prec \beta \prec 2\pi,$ then Eq.(2.6) has only an imaginary solution given by:

$$\alpha = iLog[2 - Cos\beta + \sqrt{3 - 4Cos\beta + Cos^2\beta}]. \tag{2.7}$$

For the case $l = 0,$ we may write:

$$Q_{0,m} = C\int_0^{2\pi} d\beta F(\beta)\exp(im\beta)[4 - 2\exp(i\alpha) - 2Cos\beta];$$

$$= 2C\int_0^{2\pi} d\beta F(\beta)\exp(im\beta)[Cos\alpha - \exp(i\alpha)];$$

$$= -2iC\int_0^{2\pi} d\beta F(\beta) Sin\alpha \exp(im\beta). \tag{2.8}$$

The above charges may be construed as the coefficients of the Fourier series

$$-2iF(\beta)Sin\alpha = \frac{1}{2\pi}\sum_{m=-\infty}^{\infty} Q_{0,m}\exp(-im\beta). \tag{2.9}$$

Now, take $Q_{0,m} = \delta_{0m}$; i.e. $Q_{0,0} = 1,$ and $Q_{0,m} = 0.$ This situation correspond to the case where a charge $Q$ enters at the node *(0,0)* and leaves at infinity. Note that no charges leave the lattice at any other finite node. So, with this choice

$$F(\beta) = \frac{i}{4\pi Sin\alpha}. \tag{2.10}$$

Thus, we can write:

$$V_{l,m} = \frac{i}{4\pi} \int_0^{2\pi} \frac{d\beta}{Sin\alpha} \exp(i|l|\alpha + im\beta). \qquad (2.11)$$

One may ask. What is the capacitance between the origin and the site (l,m)?. First of all, it is clear that the potential difference between these two sites is $(V_{0,0} - V_{l,m})$. Now, imagine that a charge of one micro *Coulomb* enters the network at the node *(l,m)* instead of *(0,0)*, allowing it to leave at infinity. The new potential at *(l,m)* will now be what we called $V_{0,0}$ and the new potential at *(0,0)* will be, by symmetry, what we called $V_{l,m}$. Thus, the new potential difference between the origin and *(l,m)* is just minus the previous one.

If we choose to imagine that one micro *Coulomb* leaves the node *(l,m)*, so that all the potential difference will be reversed in sign. Therefore, the new potential difference between the origin and the node *(l,m)* is again given by $(V_{0,0} - V_{l,m})$.

Exploiting the linearity of Ohm's law and superpose all the charges and potentials appertaining to the configuration in which one micro *Coulomb* enters at *(0,0)* and leaves at *(l,m)*, one can write:

$$2[V_{0,0} - V_{l,m}]C_{l,m} = 1. \qquad (2.12)$$

Or,

$$C_{l,m} = \frac{1}{2[V_{0,0} - V_{l,m}]}. \qquad (2.13)$$

Thus, $C_{l,m}$ can be written as:

$$C_{l,m} = \frac{1}{\frac{i}{2\pi} \int_0^{2\pi} \frac{d\beta}{Sin\alpha}[1 - \exp(i|l|\alpha + im\beta)]}. \qquad (2.14)$$

It is obvious that $C_{l,m} = C_{m,l}$ due to the symmetry of the lattice. Finally, we may transform Eq. (2.14) into the manifestly real form:

$$C_{l,m} = \frac{1}{\frac{1}{\pi} \int_0^{\pi} \frac{d\beta}{Sinh|\alpha|}[1 - \exp(-|l||\alpha|)Cosm\beta]}. \qquad (2.15)$$

As $l \to \infty$, $C_{\infty,m} \to \dfrac{1}{\dfrac{1}{\pi}\int_0^\pi \dfrac{d\beta}{Sinh|\alpha|}} \to 0$.

Using Eq. (2.15), one can calculate the capacitance $C_{l,m}$ by means of Mathematica. Table 1 below shows some calculated values. Similar results were obtained by us[22] using similar approach, and by using LGF approach[7].

## 3. Infinite SC Network

The above method can be generalized to an infinite three-dimensional SC network of identical capacitors. Here we consider three indices and a charge entering the site *(l,m,n)* is related to the potentials by:

$$Q_{l,m,n} = (6V_{l,m,n} - V_{l+1,m,n} - V_{l-1,m,n} - V_{ll,m+1,n} - V_{l,m-1,n} - V_{l,m,n+1} - V_{l,m,n-1})C. \qquad (3.1)$$

Choose $V_{l,m,n}$ to be given as:

$$V_{l,m,n} = \int_0^{2\pi} d\beta \int_0^{2\pi} d\gamma F(\beta,\gamma) v_{l,m,n}(\beta,\gamma). \qquad (3.2)$$

Where $v_{l,m,n}(\beta,\gamma) = \exp(i|l|\alpha + im\beta + in\gamma)$, and with $Cos\alpha + Cos\beta + Cos\gamma = 3$. i.e.; $\alpha = Cos^{-1}(3 - Cos\beta - Cos\gamma)$.
For $l \neq 0$, one can easily show that:
$$Q_{l,m,n} = 2\int_0^{2\pi} d\beta \int_0^{2\pi} d\gamma F(\beta,\gamma) \exp(i|l|\alpha + im\beta + in\gamma) x [3 - Cos\alpha - Cos\beta - Cos\gamma]. \qquad (3.3)$$

whereas:

$$Q_{0,m,n} = -2i\int_0^{2\pi} d\beta \int_0^{2\pi} d\gamma F(\beta,\gamma) Sin\alpha Cosm\beta Cosn\gamma. \qquad (3.4)$$

The inverse of this double Fourier series gives:
$$-2iF(\beta,\gamma)Sin\alpha = \dfrac{4}{\pi^2} \sum_{m=-\infty}^{\infty} \sum_{n=-\infty}^{\infty} Q_{0,m,n} \exp(-im\beta)\exp(-in\gamma). \qquad (3.5)$$

Let us choose $Q_{0,0,0} = 1$, and $Q_{0,m,n} = 0$, unless both *m* and *n* vanish. This implies that:

$$F(\beta,\gamma) = \frac{i}{8\pi^2 Sin\alpha}. \tag{3.6}$$

Substitute Eq. (3.6) into Eq. (3.2), it yields the potential $V_{l,m,n}$. That is,

$$V_{l,m,n} = \int_0^{2\pi} d\beta \int_0^{2\pi} d\gamma \frac{i}{8\pi^2 Sin\alpha} v_{l,m,n}(\beta,\gamma). \tag{3.7}$$

As in section 2, we can compute the capacitance $C_{l,m,n}$ by assuming that a charge of one micro *Coulomb* enters the origin and leaves the lattice site *(l,m,n)*. Thus;

$$C_{l,m,n} = \frac{1}{\frac{i}{4\pi^2} \int_0^{2\pi} \int_0^{2\pi} \frac{d\beta d\gamma}{Sin\alpha}[1 - \exp(i|l|\alpha) + im\beta + in\gamma]}. \tag{3.8}$$

Again, this expression is symmetric under any permutation of the indices. A manifestly real form of Eq. (3.8) is:

$$C_{l,m,n} = \frac{1}{\frac{1}{\pi^2} \int_0^{\pi} \int_0^{\pi} \frac{d\beta d\gamma}{Sinh|\alpha|}[1 - \exp(-|l||\alpha|)Cosm\beta Cosn\gamma]}. \tag{3.9}$$

As $l \to \infty$, $C_{\infty,m,n} \to \dfrac{1}{\frac{1}{\pi^2}\int_0^\pi \int_0^\pi \frac{d\beta\, d\gamma}{Sinh|\alpha|}} \to 0$.

Using Eq. (3.9), we can calculate $C_{l,m,n}$ by means of Mathematica. Table 2 below shows some calculated values.

This method can be straight forwardly generalized to four and more dimensions. In a *(d+1)* dimensional hypercubic lattice, the capacitance between the origin and the lattice site $(m_1, m_2,..., m_d)$ is

$$C_{m_1,...,m_d} = \frac{1}{\frac{i}{(2\pi)^d}\int_0^{2\pi}...\int_0^{2\pi} \frac{d\beta_1...d\beta_d}{Sin\alpha}[1 - \exp(i|m_1|\alpha) + im_2\beta_2 + ... + im_d\beta_d]}. \tag{3.10}$$

Where $Cos\alpha + Cos\beta_1 + ... + Cos\beta_d = d$.

As a further generalization, consider the SC network with different capacitances in the three directions: for example say, 1 micro Farad along the x-direction, $\dfrac{1}{p}$ micro Farad along the y-direction and $\dfrac{1}{q}$ micro Farad along the z-direction. In this case the charge entering the lattice site $(l,m,n)$ is given by:

$$Q_{l,m,n} = 2(1+p+q)V_{l,m,n} - V_{l+1,m,n} - pV_{l,m+1,n} - pV_{l,m-1,n} - qV_{l,m,n+1} - qV_{l,m,n-1}. \quad (3.11)$$

Therefore the capacitance $C_{l,m,n}$ is still given by Eq. (3.10), but with

$$\alpha = Cos^{-1}(1+p+q-pCos\beta-qCos\gamma). \quad (3.12)$$

Now, with $p=q=1$ we recover the symmetric SC lattice, while for $p=1$ and $q=0$ give the square lattice discussed in section 2 above. Finally, for $p \neq 1$ and $q=0$ we got the "rectangular" lattice (i.e. a square lattice with unequal capacitances in the two coordinate directions).

## 4. Results and Discussion

This work is divided into two main parts. In part one, the capacitance between the site *(0,0)* and the site *(l,m)* in an infinite square grid consisting of identical capacitors is calculated using the superposition of charge distribution. The capacitance $C_{l,m}$ is expressed in an integral form which is evaluated analytically and numerically (Table 1). While in part two, the capacitance between the site *(0,0,0)* and the site *(l,m,n)* in an infinite SC grid consisting of identical capacitors is also calculated using the superposition of charge distribution. The capacitance $C_{l,m,n}$ is expressed in an integral form as the infinite square grid which is evaluated analytically and numerically (Table 2).

In Figs. 1 and 2 the capacitance is plotted against the site *(l,m)*. Figure 1 shows the capacitance of the infinite square grid as a function of *l* and *m* along [10] direction and Fig. 2 shows the capacitance of the infinite square grid as a function of *l* and *m* along [11] direction. Inversion symmetry is present in both figures around the origin.

In Figs 3 and 4 the capacitance is plotted against the site *(l,m,n)*. Figure 3 shows the capacitance of the infinite square grid as a function of *l*, *m*, and *n* along [100] direction and Fig. 4 shows the capacitance of the infinite square grid as a function of *l*, *m*, and *n* along [111] direction. One can notice from these two figures that there are inversion symmetries about the origin.

The asymptotic form of equations (2.15) and (3.9) corresponding to the identical capacitors of infinite square lattice and infinite simple cubic lattice, respectively leads to zero as l goes to infinity(see Figs. 1 and 3).

An investigation of infinite complicated lattices and of lattices with missing capacitor (bond) is in progress.

## Table Captions

**Table 1:** Numerical values of $C_{l,m}$ in units of $C$ for an infinite square grid.

**Table 2:** Numerical values of $C_{l,m,n}$ in units of $C$ for an infinite SC grid.

## Table 1

| $l, m$ | $C_{l,m}$ |
|---|---|
| 0 | ∞ |
| 1,0 | 2 |
| 2,0 | 1.37597 |
| 3,0 | 1.16203 |
| 4,0 | 1.04823 |
| 5,0 | 0.974844 |
| 1,1 | 1.5708 |
| 2,1 | 1.29326 |
| 3,1 | 1.13539 |
| 4,1 | 1.03649 |
| 5,1 | 0.968523 |
| 2,2 | 1.1781 |
| 3,2 | 1.08177 |
| 4,2 | 1.00814 |
| 5,2 | 0.951831 |
| 3,3 | 1.02443 |
| 4,3 | 0.972869 |
| 5,3 | 0.929041 |
| 4,4 | 0.937123 |
| 5,4 | 0.90391 |
| 5,5 | 0.878865 |

**Table 2**

| $l,m,n$ | $C_{l,m,n}$ | $l,m,n$ | $C_{l,m,n}$ | $l,m,n$ | $C_{l,m,n}$ |
|---|---|---|---|---|---|
| 0 | ∞ | 410 | 2.144 | 531 | 2.08958 |
| 100 | 3.000003 | 411 | 2.138018 | 532 | 2.084667 |
| 110 | 2.531139 | 420 | 2.12867 | 533 | 2.077909 |
| 111 | 2.3906 | 421 | 2.124356 | 540 | 2.080503 |
| 200 | 2.382751 | 422 | 2.113825 | 541 | 2.079348 |
| 210 | 2.306284 | 430 | 2.111192 | 542 | 2.075559 |
| 211 | 2.264847 | 431 | 2.108277 | 543 | 2.070342 |
| 220 | 2.225432 | 432 | 2.100721 | 544 | 2.064235 |
| 221 | 2.206804 | 433 | 2.09079 | 550 | 2.070179 |
| 222 | 2.173162 | 440 | 2.094776 | 551 | 2.069768 |
| 300 | 2.220392 | 441 | 2.092865 | 552 | 2.066637 |
| 310 | 2.200632 | 442 | 2.087565 | 553 | 2.062804 |
| 311 | 2.186289 | 443 | 2.0803 | 554 | 2.057948 |
| 320 | 2.167735 | 444 | 2.072238 | 555 | 2.05287 |
| 321 | 2.159146 | 500 | 2.11299 | 600 | 2.088777 |
| 322 | 2.14053 | 510 | 2.109767 | 610 | 2.087086 |
| 330 | 2.136601 | 511 | 2.106833 | 633 | 2.069498 |
| 331 | 2.131646 | 520 | 2.101692 | 644 | 2.056729 |
| 332 | 2.119735 | 521 | 2.099336 | 655 | 2.047817 |
| 333 | 2.105161 | 522 | 2.093075 | 700 | 2.071745 |
| 400 | 2.15107 | 530 | 2.091324 | 531 | 2.08958 |

**Figure Captions**

**Fig. 1:** The capacitance $C_{l,m}$ in terms of *l* and *m* for an infinite square grid along the [10] direction.

**Fig. 2:** The capacitance $C_{l,m}$ in terms of *l* and *m* for an infinite square grid along the [11] direction.

**Fig. 3:** The capacitance $C_{l,m,n}$ in terms of *l*, *m*, and *n* for an infinite SC grid along the [100] direction.

**Fig. 4:** The capacitance $C_{l,m,n}$ in terms of *l*, *m*, and *n* for an infinite SC grid along the [111] direction.

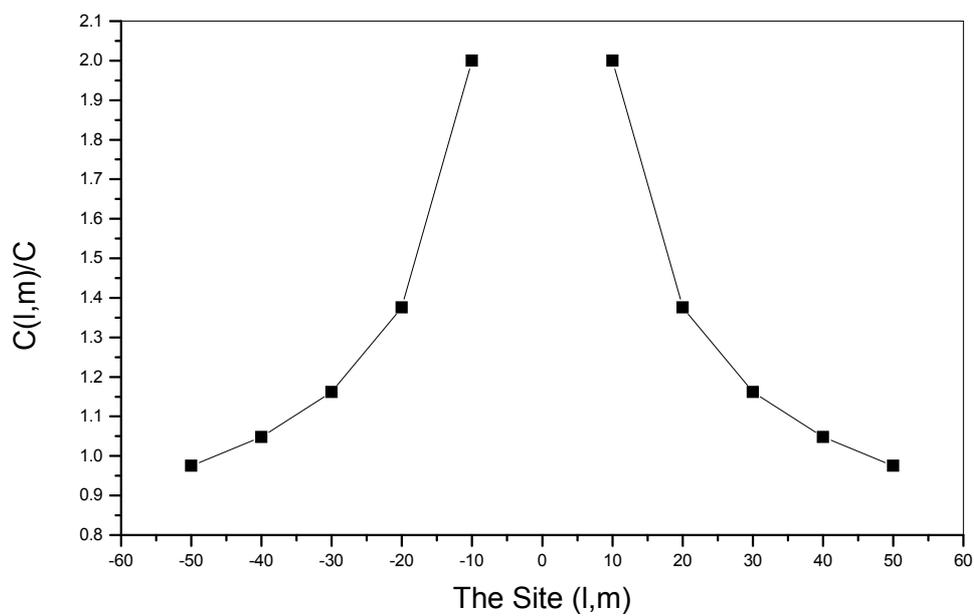

**Fig. 1**

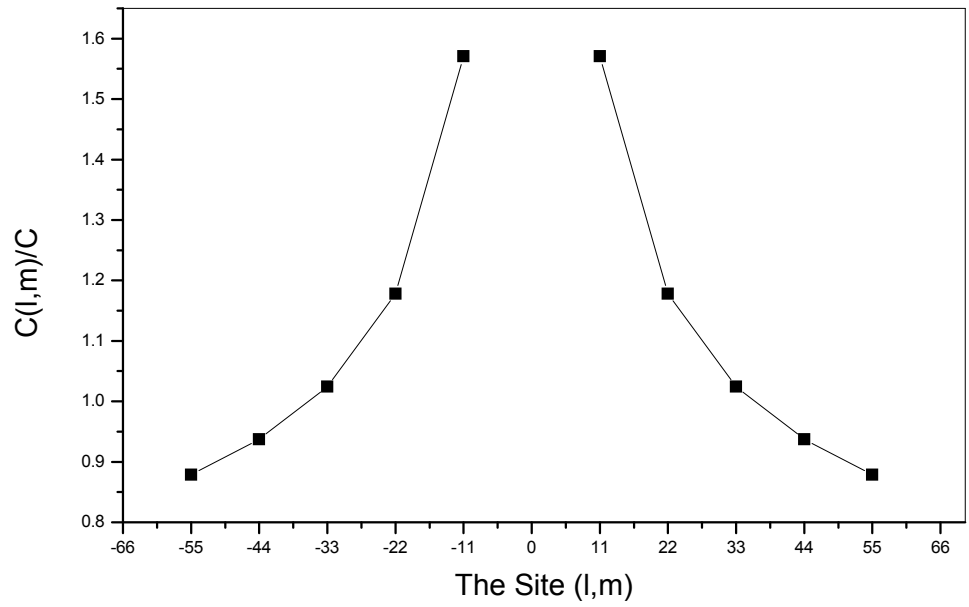

**Fig. 2**

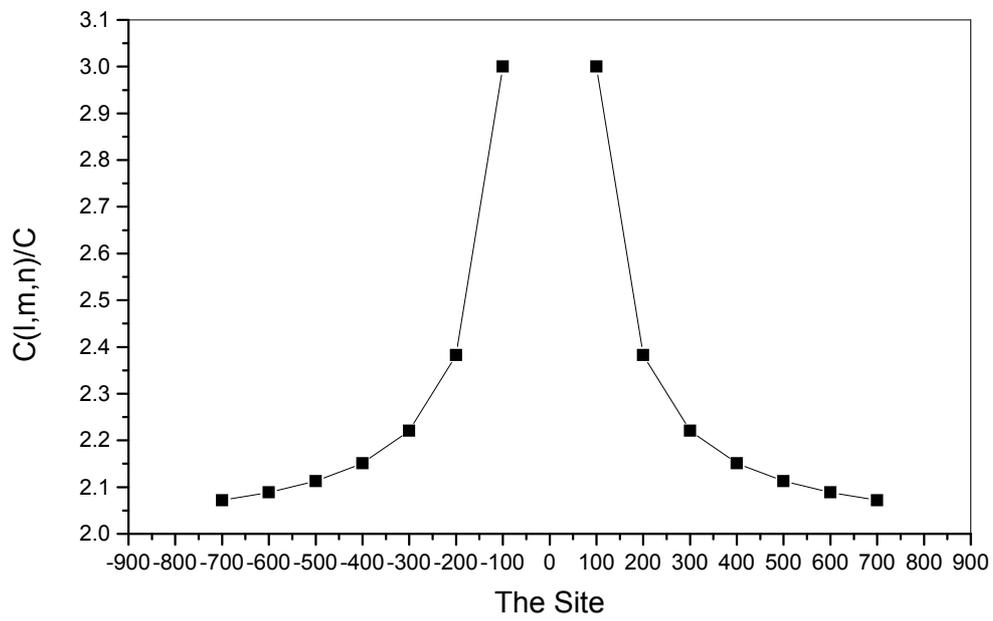

**Fig. 3**

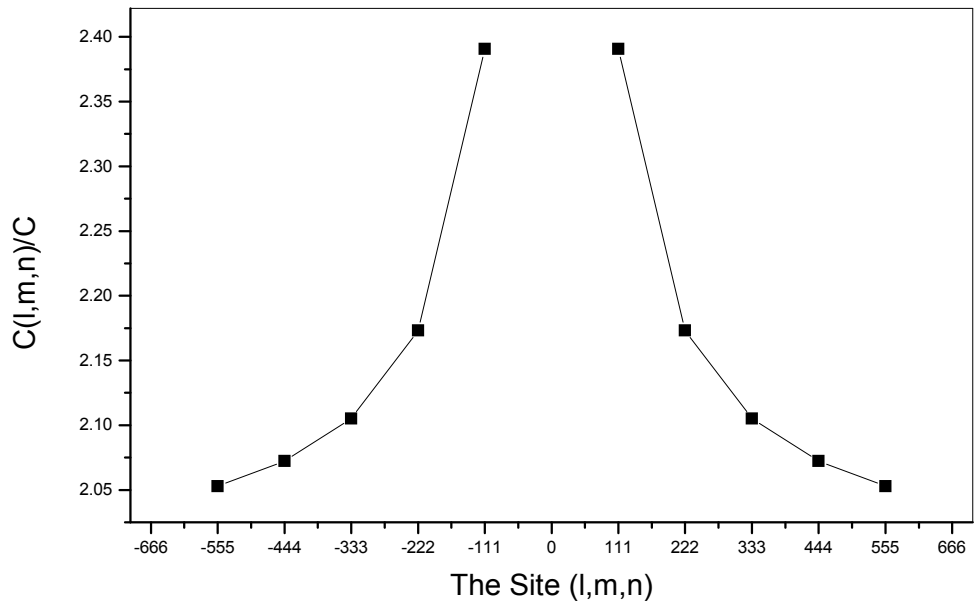

**Fig. 4**